\documentclass[journal]{IEEEtran}
\usepackage{amsmath}
\usepackage{amssymb}
\usepackage{amsfonts}
\usepackage{graphicx}
\usepackage{float}
\usepackage{subfigure}
\usepackage{booktabs}
\usepackage{multirow}
\usepackage{diagbox}
\usepackage{stfloats}
\usepackage{cite}
\usepackage{url}
\usepackage{xspace}
\usepackage{siunitx}

\newcommand{\etal}{\emph{et al.}\xspace}
\newcommand{\ie}{\emph{i.e.}\xspace}

\newcommand{\mat}[1]{\boldsymbol{#1}} % matrix

\makeatletter

\newcommand{\Rmnum}[1]{\expandafter\@slowromancap\romannumeral #1@}
\makeatother
\author{Yu~Gong,
Hongming~Shan,~\IEEEmembership{Member,~IEEE},
Yueyang~Teng$^\ast$,
Ning~Tu,
Ming~Li,
Guodong~Liang,
Ge~Wang,~\IEEEmembership{Fellow,~IEEE},
and ~Shanshan~Wang,~\IEEEmembership{Senior Member,~IEEE} % <-this % stops a space
\thanks{This work was supported by the National Natural Science Foundation of China (61871371, 81830056, 61671441), Natural Science Foundation of Liaoning Province of China (20170540321), Science and Technology Planning Project of Guangdong Province (2017B020227012, 2018B010109009), the Basic Research Program of Shenzhen (JCYJ20180507182400762), Youth Innovation Promotion Association Program of Chinese Academy of Sciences (2019351), NIH/NCI under award numbers R01CA233888 and R01CA237267, and NIH/NIBIB under award number R01EB026646. Asterisk indicates the corresponding author.}
\thanks{Y. Gong is with the College of Medicine and Biological Information Engineering, Northeastern University, Shenyang 110169, China, and with Paul C. Lauterbur Research Center for Biomedical Imaging, Shenzhen Institutes of Advanced Technology, Chinese Academy of Sciences, Shenzhen 518055, China (e-mail: gongyu0010@gmail.com).}% <-this % stops a space
\thanks{Y. Teng is with the College of Medicine and Biological Information Engineering, Northeastern University, Shenyang 110169, China, and with the Key Laboratory of Intelligent Computing in Medical Images, Ministry of Education, Shenyang 110169, China (e-mail: tengyy@bmie.neu.edu.cn).}
\thanks{H. Shan and G. Wang are with the Department of Biomedical Engineering, Rensselaer Polytechnic Institute, Troy, NY 12180 USA (e-mail: shanh@rpi.edu; wangg6@rpi.edu).}% <-this % stops a space

\thanks{N. Tu is with the PET-CT/MRI Center and Molecular Imaging Center, Wuhan University Renmin Hospital, Wuhan, 430060, China (e-mail: tuning@whu.edu.cn).}% <-this % stops a space
\thanks{M. Li and G. Liang are with the Neusoft Medical Systems Co., Ltd, Shenyang 110167, China (e-mail: ming\_li@neusoft.com; lianggd@neusoft.com).}
\thanks{S. Wang is with Paul C. Lauterbur Research Center for Biomedical Imaging, Shenzhen Institutes of Advanced Technology, Chinese Academy of Sciences, Shenzhen 518055, China (e-mail: ss.wang@siat.ac.cn).}% <-this % stops a space
}
\title{Parameter-Transferred Wasserstein Generative Adversarial Network (PT-WGAN) for Low-Dose PET Image Denoising}

\begin{document}
\maketitle
\begin{abstract} Due to the widespread use of positron emission tomography (PET) in clinical practice, the potential risk of PET-associated radiation dose to patients needs to be minimized. However, with the reduction in the radiation dose, the resultant images may suffer from noise and artifacts that compromise diagnostic performance. In this paper, we propose a parameter-transferred Wasserstein generative adversarial network (PT-WGAN) for low-dose PET image denoising. The contributions of this paper are twofold: i) a PT-WGAN framework is designed to denoise low-dose PET images without compromising structural details, and ii) a task-specific initialization based on transfer learning is developed to train PT-WGAN using trainable parameters transferred from a pretrained model, which significantly improves the training efficiency of PT-WGAN. The experimental results on clinical data show that the proposed network can suppress image noise more effectively while preserving better image fidelity than recently published state-of-the-art methods. We make our code available at \url{https://github.com/90n9-yu/PT-WGAN}.

\end{abstract}

\section{Introduction}
Positron emission tomography (PET) is an advanced imaging device in the field of nuclear medicine and plays an important role in neurology \cite{gunn2015quantitative}, oncology \cite{beyer2000combined}, and cardiology \cite{machac2005cardiac}. As a noninvasive tool, PET has many clinical applications, such as cancer screening, tumor detection, tumor staging, and monitoring of treatment responses and outcomes. For a PET scan, a patient needs to be injected with a small amount of a radioactive tracer, for example, 18F-fluorodeoxyglucose (FDG). The positrons emitted from it are annihilated almost immediately in the patient body to emit paired $\gamma$ photons. The use of less radioactive tracer means less cost and less risk, making PET scans safer for patients and staff. In recent years, researchers have attempted to reduce the dose of radioactive tracers used in PET scans \cite{fahey2019novel}. PET dose reduction follows the well-known guiding principle of ALARA (as low as reasonably achievable) \cite{brenner2007computed}. Due to various physical degradation factors and low coincident photon counts detected, reducing the dose of radioactive tracers significantly compromises the final imaging quality. Therefore, advanced imaging technology \cite{kwon2014signal} and imaging algorithms \cite{Reader_1998} are desirable for denoising low-dose PET images.

Existing PET image denoising algorithms can be divided into two categories: iterative reconstruction algorithms and image postprocessing algorithms. The iterative reconstruction algorithm combines the statistical model of data with a regularization term over an image to suppress the noise in a reconstructed image. For instance, Wang \textit{et al.} \cite{wang2012penalized} proposed a patch-based regularization to preserve the edge of an image in the iterative process to obtain high-quality PET images. Ehrhardt \textit{et al.} \cite{ehrhardt2018faster} proposed using randomized optimization for PET reconstruction aided by a large class of nonsmooth priors. Reader \textit{et al.} \cite{Reader_1998} proposed a fast accurate iterative reconstruction (FAIR) method based on the expectation maximization-maximum likelihood (EM-ML) technique. Xu \textit{et al.} \cite{Xu_2017} proposed utilizing total variation (TV) penalized with a uniform interval constraint algorithm for medical image denoising. With the development of multimodality imaging technology, using anatomical prior information to guide PET reconstruction has become a popular research area. For example, Bland \textit{et al.} \cite{bland2017mr} proposed a magnetic resonance (MR)-guided kernel method for low-dose PET reconstruction. Song \textit{et al.} \cite{song2019pet} proposed an image deblurring and superresolution framework for PET using anatomical guidance provided by high-resolution MR images. The iterative reconstruction algorithm shows an excellent denoising ability, but there are limitations as well; for example, the iterative reconstruction algorithm is computationally intensive and may induce additional artifacts. On the other hand, image postprocessing after reconstruction is computationally efficient compared to iterative reconstruction. Over the past years, many excellent image postprocessing algorithms have been proposed, such as image-guided filtering for dynamic sinogram denoising \cite{hashimoto2018denoising}, nonlocal means filtering \cite{zhang2017applications} and block-matching 3D filtering \cite{hasan2018denoising}. Although denoising through image postprocessing may substantially improve the image quality, oversmoothing and residual artifacts are often observed in the denoised image.

Recently, deep learning has achieved extraordinary results in the field of medical images, such as image segmentation \cite{zhao2019automatic,guo2019deep,yang2019clci,hatt2018first}, image reconstruction \cite{wang2016accelerating,wang2020deepcomplexmri,chen2019model}, and image denoising \cite{xiao2019stir,yi2018sharpness,kadimesetty2018convolutional}. Generally, the statistical characteristic of noise in medical images is complicated and difficult to model. Deep learning can solve this problem well due to its powerful learning ability driven by big data. Therefore, deep learning-based medical image noise reduction methods have led to state-of-the-art results, clearly outperforming traditional methods, as evidenced in \cite{shan2019competitive}. For PET image denoising, the deep learning-based method can be divided into two categories. One category is only low-dose PET images as input (PET-only model) \cite{gong2018pet,wang20183d}. Gong \textit{et al.} \cite{gong2018pet} proposed using a deep neural network for PET denoising. The proposed network fine-tuned the last few layers using real datasets after the pretraining by simulation datasets to overcome the data dependence. Wang \textit{et al.} \cite{wang20183d} proposed using the generative adversarial network (GAN) \cite{goodfellow2014generative} to estimate high-quality normal-dose PET images from the corresponding low-dose PET images. The other category is PET images with additional MR images as input (PET-MR model) \cite{xiang2017deep,cui2019pet,chen2019ultra}. Xiang \textit{et al.} \cite{xiang2017deep} proposed a network with the combination of low-dose PET images and T1-weighted MR images as input to estimate normal-dose PET images. Chen \textit{et al.} \cite{chen2019ultra} proposed using multicontrast MR images to exploit anatomical information for ultralow-dose PET image denoising.

Currently, most of denoising methods are in the 2D domain, which means that only 2D information is utilized. Given that PET is a 3D imaging technology, experienced radiologists usually observe many adjacent slices along the $z$-dimension for clinical analysis. It is advisable to target 3D information for low-dose PET image denoising. Shan \textit{et al.} \cite{shan20183} proposed a 3D convolutional encoder-decoder network for low-dose CT transferred from a 2D trained network to achieve significant denoising performance. The 3D network explores 3D contextual information, but it suffers from high computational resource demand. The 2D network is computationally efficient. However, it ignores the 3D information of PET images.

In this paper, we propose a hybrid 2D and 3D parameter-transferred Wasserstein generative adversarial network (PT-WGAN) for low-dose PET image denoising. It is a tradeoff between computational effectiveness and denoising performance. The generator of PT-WGAN is designed for noise reduction, and consists of 2D and 3D encoder-decoder structures. There are typical convolution operators and deconvolution operators in the encoder and decoder respectively. In the generator, we first use 3D convolution operators and then use 2D convolution operators to combine features in the 3D and 2D domains, which bridges the gap between the 3D and 2D feature spaces and saves computational resources. Convolution operators are used to capture the abstraction of image contents while reducing noise and corruption. Deconvolution operators are then used to recover image details. To reduce the training difficulty of WGAN, we propose a task-specific initialization based on transfer learning to stabilize the training process. First, we use the mean squared error (MSE) as the loss function to train the generator of the proposed network separately. In the next training phase, we propose leveraging the task-specific initialization instead of using the Xavier initialization \cite{glorot2010understanding}. Specifically, we use the parameters of the pretrained model to initialize the generator. The experimental results on clinical images show that the proposed network can suppress more noise while preserving more details than three state-of-the-art methods.

The contributions of this paper are as follows:
\begin{enumerate}
\item[1)] A PT-WGAN framework with a hybrid structure of 2D and 3D encoder-decoder networks is designed for low-dose PET image denoising. It maps the distribution of low-dose PET image volumes to that of normal-dose image volumes to reduce noise and retain as many faithful details as possible. The proposed network achieves a comparable denoising performance to that of the pure 3D network, while the former uses only 50\% of the number of parameters of the latter. As a result, the significantly reduced computational cost is beneficial in clinical applications.

\item[2)] A task-specific initialization based on transfer learning is proposed in the training process of PT-WGAN. It reduces the training computational cost and difficulty of the proposed network.
\end{enumerate}

\section{METHODS}
\subsection{Low-dose PET Denoising}
Let $\mat{V}_{\mathrm{LD}}$ denote the low-dose PET image volume and $\mat{V}_{\mathrm{ND}}$ denote the corresponding normal-dose image volume. The task of the denoising process is to seek a function $f$ that maps the low-dose PET image volume to the normal-dose image volume. 
\begin{equation}
\label{Eq.1}
f:\mat{V}_{\mathrm{LD}}\rightarrow \mat{V}_{\mathrm{ND}}.
\end{equation}
If we take $\mat{V}_{\mathrm{LD}}$ as a sample from the low-dose PET image volume distribution $\mathbb{P}_{\mathrm{LD}}$ and $\mat{V}_{\mathrm{ND}}$ from the normal-dose PET image volume distribution $\mathbb{P}_{\mathrm{ND}}$, then the function $f$ can be used to map samples from $\mathbb{P}_{\mathrm{LD}}$ into a certain denoised distribution $\mathbb{P}_{\mathrm{denoised}}$. 
%Editor: Please ensure that the intended meaning has been maintained in the above edit.
We can make $\mathbb{P}_{\mathrm{denoised}}$ close to $\mathbb{P}_{\mathrm{ND}}$ by varying the function $f$. In other words, we treat the denoising process as a converter that can move one data distribution to another.

\subsection{PT-WGAN}
As shown in Eq. \eqref{Eq.1}, the main task for low-dose PET image denoising is to construct the function $f$. However, it is an ill-posed problem and motivates us to leverage the powerful nonlinear fitting ability of deep learning. Specifically, we design a network named PT-WGAN to approximate $f$.

The introduction of PT-WGAN consists of two parts: network structure and objective function.

\subsubsection{Network Structure}
\begin{figure}[!hbtp]
\centering
\includegraphics[width=1\linewidth]{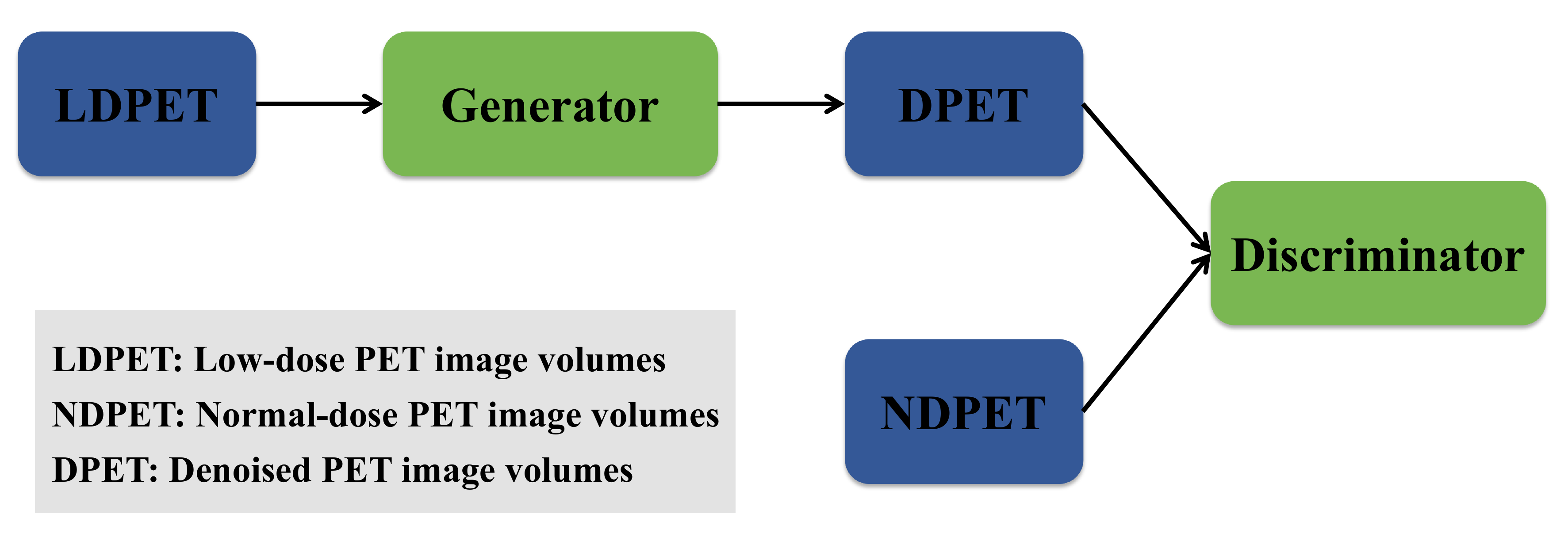}
\caption{The overall structure of the proposed PT-WGAN network}
\label{Fig.1}
\end{figure}

The overall structure of the proposed network is shown in Fig. \ref{Fig.1} and has two parts.

\begin{figure*}[!htbp]
\centering
\includegraphics[width=0.9\linewidth]{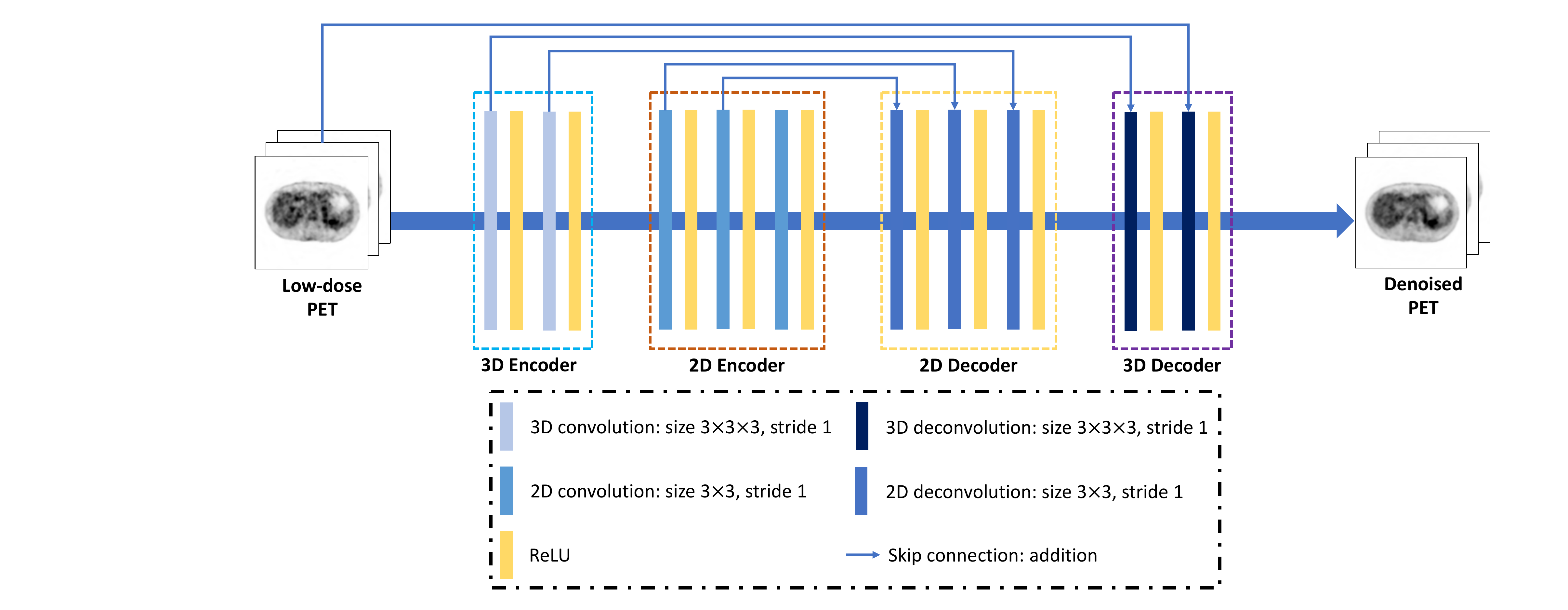}
\caption{Structure of the generator of PT-WGAN}
\label{Fig.2}
\end{figure*}

The first part is the generator. The input and the output of the generator are both image volumes. Although it has a similar shape to RED-CNN \cite{chen2017low} and RED-Net \cite{mao2016image}, the difference is that it uses hybrid 2D and 3D convolution and deconvolution operators instead of pure 2D or pure 3D convolution and deconvolution operators. 
%Editor: Abbreviations and acronyms are often defined the first time they are used within the abstract and again in the main text and then used throughout the remainder of the manuscript. Please consider adhering to this convention. The target journal may have a list of abbreviations that are considered common enough that they do not need to be defined.
The purpose of doing so is to utilize the 3D information in the PET image volume while saving computational resources. The architecture of the generator is shown in Fig. \ref{Fig.2}. There are ten layers in the generator, including two 3D convolution layers, three 2D convolution layers, three 2D deconvolution layers, and two 3D deconvolution layers. Except for the last layer, each layer consists of thirty-two kernels. The last layer consists of one kernel. The kernel size of the 3D convolution and deconvolution layers is 3$\times$3$\times$3. The kernel size of the 2D convolution and deconvolution layers is 3$\times$3. The stride of the convolution and deconvolution is a constant value of 1. To minimize the loss of information along the $z$-dimension with inferior quality, zero-padding is used in the 3D convolution and deconvolution layers. However, zero-padding is neither used in the 2D convolution layers nor the deconvolution layers for saving computational resources. The size of the output of the network is kept the same as the size of the input of the network. ReLU is used as the activation function after each layer.

The 3D convolution layers are used in the generator to extract the 3D feature maps in a PET image volume. After that, 2D convolution layers, instead of 3D convolution layers, are used to handle the extracted feature maps. This approach maintains accuracy while saving computational resources. Note that the output of the last 3D convolution layer is a 5D tensor while the input to the following 2D convolution layer is a 4D tensor. We split it along the $z$-dimension, which refers to the depth of the image, to have nine 4D tensors and send them to the following 2D convolution layer. After being processed by the 2D layers, the nine 4D tensors are concatenated along the $z$-dimension to form a 5D tensor as the input to the following 3D deconvolution layer. The noise level is reduced after each convolution layer. However, the image details may also be lost. Although the deconvolution is introduced to compensate for details, the deconvolution may not be effective in recovering details when the network becomes deeper \cite{mao2016image}. We introduce a skip connection similar to that reported in \cite{chen2017low,mao2016image}. The difference between our work and prior arts lies in that the skip connection is expanded for the 3D domain. As shown in Fig. \ref{Fig.2}, the output of the deconvolution layer is added to the output of the mirrored convolution layer. The feature maps passed by the skip connections preserve details, which helps the deconvolution layers recover a clean image volume. In addition, it is helpful for reducing the training difficulty caused by gradient diffusion.

The second part is the discriminator. There are six layers including four 3D convolution layers and two fully connected layers. Leaky ReLU is chosen as the activation function for the discriminator and is used for each layer except for the last layer. Small 3$\times$3$\times$3 kernels are used in the discriminator. The stride of the convolution layer is 2. The number of kernels and outputs of each layer is shown in Table \ref{Table.1}.

\begin{table}[!htbp]
\caption{Number of kernels and outputs of each layer of the discriminator}\label{Table.1}
\centering
\begin{tabular}{ccccccc}
\toprule
Layer & 1 & 2 & 3 & 4 & 5 & 6 \\ \midrule
\# Kernel & 64 & 128 & 256 & 512 & 1024 & 1 \\
\bottomrule
\end{tabular}
\end{table}

\subsubsection{Objective Function}
A typical WGAN consists of a pair of networks: a generator $G$ and a discriminator $D$ described above. $G$ is designed to approximate the function $f$ described in Section II-A. The difference between the original GAN \cite{goodfellow2014generative} and WGAN \cite{arjovsky2017wasserstein} is the use of the Wasserstein distance instead of the Jensen-Shannon (JS) divergence to compare the data distributions. In this paper, we use the Wasserstein distance to measure the difference between $\mathbb{P}_{\mathrm{denoised}}$ and $\mathbb{P}_{\mathrm{ND}}$. WGAN-GP \cite{gulrajani2017improved} introduced the gradient penalty to improve the tractability of WGAN. Following \cite{gulrajani2017improved}, the adversarial loss is defined as Eq. \eqref{Eq.2}:

\begin{equation}
\begin{aligned}
\label{Eq.2}
\mathcal{L}_{\mathrm{adv}} =& \underbrace{\mathbb{E}_{\mat{V}_{\mathrm{ND}} \sim \mathbb{P}_{\mathrm{ND}}}\left[D(\mat{V}_{\mathrm{ND}})\right]-\mathbb{E}_{\mat{V}_{\mathrm{LD}} \sim \mathbb{P}_{\mathrm{LD}}}\left[D\left(G(\mat{V}_{\mathrm{LD}}) \right) \right]}_{\text {Wasserstein Distance}}\\
+&\underbrace{\lambda_{gp} \mathbb{E}_{\widehat{\mat{V}}} \|\nabla_{\widehat{\mat{V}}} D(\widehat{\mat{V}}) - 1\|^2}_{\text {gradient penalty}}
\end{aligned}
\end{equation}
where $\widehat{\mat{V}}$ is a synthesis sample between real and fake samples; \ie, $\widehat{\mat{V}} = \epsilon \cdot \mat{V}_\mathrm{denoised} + (1-\epsilon)\cdot \mat{V}_\mathrm{ND} = \epsilon \cdot G(\mat{V}_\mathrm{LD}) + (1-\epsilon)\cdot \mat{V}_\mathrm{ND}$, where $\epsilon$ follows a uniform distribution $\mathcal{U}:[0,1]$. $\nabla$ is the gradient operator. $\lambda_{gp}$ denotes the weight of the gradient penalty.

Given the noise in the normal-dose image volume, the adversarial loss term tends to preserve it. However, the MSE loss term can remove it partially. Therefore, we added the MSE loss term into the training loss function. We optimized the quality of the denoised image volume by adjusting the tradeoff between the adversarial loss and MSE loss. The MSE loss is defined as:

\begin{equation}
\label{Eq.3}
\mathcal{L}_{\mathrm{MSE}}=\mathbb{E}_{(\mat{V}_{\mathrm{LD}}, \mat{V}_{\mathrm{ND}})} \| G(\mat{V}_{\mathrm{LD}})-\mat{V}_{\mathrm{ND}}\|_{F}^{2}
\end{equation}
In addition, the MSE loss helps to guarantee the pixelwise similarity between the normal-dose image volume and the denoised counterpart.

Therefore, the generator of the proposed PT-WGAN can be optimized by minimizing the following mixed loss function:

\begin{equation}
\begin{aligned}
\label{Eq.4}
\mathcal{L}_G = -\mathbb{E}_{\mat{V}_{\mathrm{LD}}}\left[D\left(G(\mat{V}_{\mathrm{LD}})\right)\right] + \lambda_m \mathcal{L}_{\mathrm{MSE}}
\end{aligned}
\end{equation}
where $\lambda_m$ denotes the tradeoff hyperparameter for the MSE loss.

Following \cite{gulrajani2017improved}, the discriminator can be optimized by minimizing the following loss function:
\begin{align}
\label{Eq.5}
\mathcal{L}_D = & \mathbb{E}_{\mat{V}_\mathrm{LD}} [D(G(\mat{V}_\mathrm{LD}))] - \mathbb{E}_{\mat{V}_\mathrm{ND}} [D(\mat{V}_\mathrm{ND})] \notag\\ 
& + \lambda_{gp} \mathbb{E}_{\widehat{\mat{V}}} \|\nabla_{\widehat{\mat{V}}} D(\widehat{\mat{V}}) - 1\|^2
\end{align}

\subsection{Task-specific Initialization}
Although the use of the Wasserstein distance and gradient penalty reduces the GAN training difficulty, the convergence problem in the training process is still not solved completely. Transfer learning is generally defined as the ability of a system to utilize knowledge learned from one task to another task that shares some common characteristics \cite{pan2009survey}. Shan \textit{et al.} \cite{shan20183} proposed to transfer a pretrained CPCE-2D model to obtain a CPCE-3D model, which improved the denoising performance and reduced the training difficulty of CPCE-3D. It is obvious that there are many common characteristics between networks that have the same structure and task but are trained by the different loss functions. The generator of the proposed network is a simple CNN network that is designed to use a low-dose PET image volume to generate a denoised image volume that is as close to the corresponding normal-dose image volume as possible, which can be trained by some common loss functions, such as the $L_{1}$-norm, MSE, and perceptual loss. However, if we use the WGAN framework, the generator can be trained by using the loss function defined as Eq. \eqref{Eq.2}. Some relations between the training process of WGANs and CNNs can be used to improve the convergence ability of the WGAN. In addition, the training process of a single CNN network is simpler than that of WGAN. Therefore, we propose a task-specific initialization based on transfer learning to train the generator without the WGAN framework from the pretrained model. It improves the efficiency of the pretrained process and is also different from the work of Shan \textit{et al.} \cite{shan20183}.

First, we tried three loss functions, including the MSE loss, structural similarity index matrix (SSIM) loss, and perceptual loss, to train the generator individually. The MSE loss is defined in Eq. \eqref{Eq.3}. The SSIM loss is defined as the average of the slicewise SSIM value calculated with the denoised image volume and the corresponding normal-dose image volume, as shown in Eq. \eqref{Eq.6}.

\begin{align}
\label{Eq.6}
\mathcal{L}_\mathrm{SSIM}= \frac{1}{N}\sum_{i=1}^{N}\left\{1- SSIM(\mat{V}_\mathrm{denoised}^{(i)}, \mat{V}_\mathrm{ND}^{(i)})\right\}&
\end{align}
where $N$ denotes the number of slices in the image volume, $\mat{V}_\mathrm{denoised}$ denotes the denoised image volume, $\mat{V}_{*}^{(i)}$ denotes the $i$-th slice from the corresponding image volume, and $SSIM$ denotes the calculation operator of SSIM, as stated in \cite{wang2004image}.

The perceptual loss is defined as the average of the slicewise perceptual similarity value calculated with the denoised image volume and the corresponding normal-dose image volume, as shown in Eq. \eqref{Eq.7}.

\begin{align}
\label{Eq.7}
\mathcal{L}_{\mathrm{per}}&=\frac{1}{N}\sum_{i=1}^{N} \| \phi(\mat{V}_\mathrm{denoised}^{(i)})-\phi(\mat{V}_\mathrm{ND}^{(i)})\|_{F}^{2}
\end{align}
where $\|\cdot\|_{F}$ represents the Frobenius norm, and $\phi(\cdot)$ denotes the feature extractor; the VGG-19 model pretrained on ImageNet has usually been used to extract perceptual features. There are sixteen convolutional layers in VGG-19, and we chose the output of the 16th convolutional layer as the extracted perceptual feature. The images were fed into VGG-19 after being normalized by a fixed mean and standard deviation for extracting features.

Then, the parameters of the trained generator were used instead of the Xavier initialization to initialize the generator in the PT-WGAN joint training process \cite{glorot2010understanding}. This kind of task-specific initialization is efficient since the training difficulty of CNN is rather low, and it provides a good starting point to prevent PT-WGAN from falling into mode collapse. This improved the training stability of the proposed network.

\subsection{Experimental Setting}
\subsubsection{Data}
Using a Neusoft NeuWise EWN ToF PET/CT scanner, nine torso PET scan datasets were acquired from nine patients, respectively, 90-110 mins after injection of fluorodeoxyglucose (F-18) in 0.15 mCi/kg. 
%Editor:  Journals often require both the manufacturer's name and location for specialized equipment and software. Please consider adding this information based on the journal’s guidelines. 
Informed consent was obtained from patients in compliance with the Institutional Review Board approval. We reconstructed the complete raw PET images from the normal-dose scans and the corresponding low-dose PET images from the 20\% uniform undersampled low-dose PET scans. Both the normal-dose and low-dose images were reconstructed using regular ordinary Poisson OSEM (OP-OSEM) with correction using the vendor’s default algorithm. The number of iterations used in OP-OSEM was set as 2, and the number of subsets as 21. Each PET volume consists of 87 contiguous slices of $256 \times 256$ pixels. The axial FOV is $165.039 \ \text{mm}$. The voxels are $2.734 \times 2.734 \times 1.897 \ {mm}^{3}$.

For training purposes, 169K volume patches of size $9 \times64 \times 64$ were randomly extracted from seven scans of patients who were randomly selected from this dataset. The remaining two patients were used to test the denoising performance of the trained models.

\subsubsection{Hyperparameters}
Similar to the training process of other GANs, we separately trained the generator $G$ and discriminator $D$ by fixing one and updating the other. The numbers of steps for training the discriminator and generator were 4:1. We used the Adam algorithm \cite{kingma2014adam} to optimize the proposed network. For the training from scratch, the hyperparameters of the Adam algorithm were set as $\alpha=1.0\times 10^{-4}$, $\beta_1=0.9$ and $\beta_2=0.999$. For transfer learning, the hyperparameters of the Adam algorithm were set as $\alpha=1.0\times 10^{-5}$, $\beta_1=0.9$ and $\beta_2=0.999$. The size of the mini-batch was 80. The number of epochs used for direct training was 40. After transfer learning, the network was trained in only 10 epochs. The weight of the gradient penalty $\lambda_{gp}$ was fixed at 10, as suggested in \cite{gulrajani2017improved}. The networks were implemented in Python 3.6 with TensorFlow 1.4 \cite{abadi2016tensorflow}. A NVIDIA GeForce RTX 2080Ti GPU was used.

To determine the weighting parameter $\lambda_m$ for the MSE loss term in the objective function, we selected the parameter $\lambda_m$ from $\{0, 10^{0}, 10^{1}, 10^{2}, 10^{3},10^{4}, 10^{5}, 10^{6}, 10^{7}, 10^{8}, 10^{9}, \infty \}$. Note that parameter $\lambda_m=0$ ($\lambda_m=\infty$) indicated that the denoising model was only optimized with the adversarial loss (MSE loss); otherwise, the denoising was optimized by balancing these two losses. As this study is designed to meet clinical needs, the radiologists’ evaluation was used as the reference. We selected $\lambda_m=10^{7}$ based on their feedback, which was used in the following experiments.

\subsection{Quantitative Evaluation Metric}
Normalized root mean square error (NRMSE), peak signal-to-noise ratio (PSNR), Riesz transform-based feature similarity (RFSIM) index \cite{zhang2010rfsim}, and visual information fidelity (VIF) \cite{sheikh2006image} were chosen for evaluating the quality of the denoised image. 

NRMSE is the root mean square error (RMSE) divided by the range of the reference image, where RMSE is defined as the square root of the MSE. The value of NRMSE is often expressed as a percentage, where lower values indicate less residual variance.

PSNR is a commonly used fidelity measure, which is related to the MSE and can be expressed as the ratio between the square of the maximum value $\mathrm{MAX}_{I}^{2}$ and the MSE value.

RFSIM proposed by Zhang \etal \cite{zhang2010rfsim} is based on the fact that the human vision system (HVS) perceives an image mainly according to its low-level features at key locations, such as edges, zero-crossings, corners, and lines. In other words, most of the crucial information for the HVS to interpret the scene is conveyed by a few key image points with salient features. Since the $1^{st}$-order and $2^{nd}$-order Riesz transform can efficiently extract several types of image low-level features in a unified theoretical framework, RFSIM is computed by comparing Riesz transform features at key locations between the reference image and the denoised image. Concerning the key locations, they are marked by a mask formed by the Canny operator (without a thinning operation).

VIF proposed by Sheikh \etal \cite{sheikh2006image} is an image quality assessment method based on the information fidelity measurement. It has been demonstrated that the visual quality of images is strongly related to relative information presented in the images. Specifically, the information of the reference images passes through the HVS channel to a human observer in the absence of any distortions. It is assumed that the information of the reference images has passed through another "distortion channel" before entering the HVS as the information of the distorted images. VIF measurement is derived from the quantification of two mutual information quantities. One is the mutual information between the input and the output of the HVS channel without the distortion channel. The other is the mutual information between the input to the distortion channel and the output of the HVS channel.

Compared with NRMSE and PSNR, RFSIM and VIF consider the correlation between the image quality and human visual perception. Theoretically, the denoised image with a lower NRMSE, a higher PSNR, a higher RFSIM, and a higher VIF presents higher quality.

\section{EXPERIMENTS}
\subsection{Ablation Experiment}

\begin{table*}[!htbp]
\caption{The details and the quantitative metrics of denoising results of the models involved in the ablation experiment. For each metric, we mark the best in bold.}\label{Table.2}
\centering
\begin{tabular}{cccccccc}
\toprule
& Structure & Loss Function & Training Method & PSNR & NRMSE (\%) & RFSIM & VIF \\ \midrule
Low-dose & - & - & - & 46.148 & 2.916 & 0.541 & 0.555 \\
Pure 2D network & 2D & $\mathcal{L}_{\mathrm{MSE}}$ & Direct & 54.551 & 1.126 & 0.745 & 0.676 \\
Pure 3D network & 3D & $\mathcal{L}_{\mathrm{MSE}}$ & Direct & 55.171 & 1.035 & 0.760 & 0.724 \\
Hybrid 2D and 3D network & 2D\&3D & $\mathcal{L}_{\mathrm{MSE}}$ & Direct & \textbf{55.348} & \textbf{1.011} & 0.761 & 0.723 \\
WGAN & 2D\&3D & $\mathcal{L}_{\mathrm{adv}}$ & Direct & 51.023 & 1.705 & 0.667 & 0.722 \\
WGAN (MSE) & 2D\&3D & $\mathcal{L}_{\mathrm{adv}}+\mathcal{L}_{\mathrm{MSE}}$ & Direct & 54.961 & 1.062 & 0.764 & \textbf{0.730} \\
PT-WGAN (MSE) & 2D\&3D & $\mathcal{L}_{\mathrm{adv}}+\mathcal{L}_{\mathrm{MSE}}$ & Transfer & 55.048 & 1.048 & \textbf{0.776} & \textbf{0.730} \\ 
PT-WGAN (SSIM) & 2D\&3D & $\mathcal{L}_{\mathrm{adv}}+\mathcal{L}_{\mathrm{MSE}}$ & Transfer & 54.919 & 1.067 & 0.758 & 0.724 \\ 
PT-WGAN (Perceptual) & 2D\&3D & $\mathcal{L}_{\mathrm{adv}}+\mathcal{L}_{\mathrm{MSE}}$ & Transfer & 54.954 & 1.061 & 0.760 & 0.726 \\ \bottomrule
\end{tabular}
\end{table*}

The ablation experiment was designed to analyze the performance of the low-dose PET image volume denoising network from the following three aspects: the network structure, the training loss function, and the training method. The details and the quantitative metrics of the denoising results of the models involved in the ablation experiment are shown in Table \ref{Table.2}. 

\subsubsection{Network structure}
%Editor: Please use a consistent capitalization and punctuation format for section headings throughout the manuscript. Some journals request a specific style, so please review the journal's guidelines.
A PET image is a full 3D volume composed of several 2D slices. The 3D network can capture 3D contexts effectively. However, the 3D network is arguably both memory-consuming and less stable \cite{ni2019elastic}. In addition, the information in the 2D domain is as important as the information in the 3D domain. The 2D network is better than the 3D network in utilizing the information in the 2D domain. In this experiment, we used the identical data and loss function (MSE) to train the pure 2D network, the pure 3D network, and the hybrid 2D and 3D network. The structure of the proposed hybrid 2D and 3D network is similar to that of the pure 2D network and the pure 3D network; for example, it includes mixed convolution and deconvolution operators and skip connections between the corresponding layers. The difference is the use of hybrid 2D and 3D convolutional operators instead of pure 2D or pure 3D convolutional operators. For fairness, we made the parameters of the pure 2D network and the pure 3D network equal by adjusting the number of kernels used in the convolutional layer in the pure 2D network to forty-eight. The numbers of parameters of the pure 2D network and the pure 3D network were approximately 220,000. The number of parameters of the hybrid 2D and 3D network was approximately 110,000.

The quantitative analysis results of the denoised image on the testing set and the details of the networks described above are shown in Table \ref{Table.2}. They are labeled as ‘Pure 2D network’, ‘Pure 3D network’, and ‘Hybrid 2D and 3D network’. As shown in Table \ref{Table.2}, the metrics of all three networks' denoised images were improved compared with that of the low-dose images. Moreover, the pure 3D network outperformed the pure 2D network. The denoising performance gap between the pure 3D network and the hybrid 2D and 3D network was small; however, the number of parameters of the hybrid 2D and 3D network was half that of the pure 3D network, indicating the proposed hybrid 2D and 3D structure was highly effective.

\begin{figure}[!htbp]
\centering
\includegraphics[width=1\linewidth]{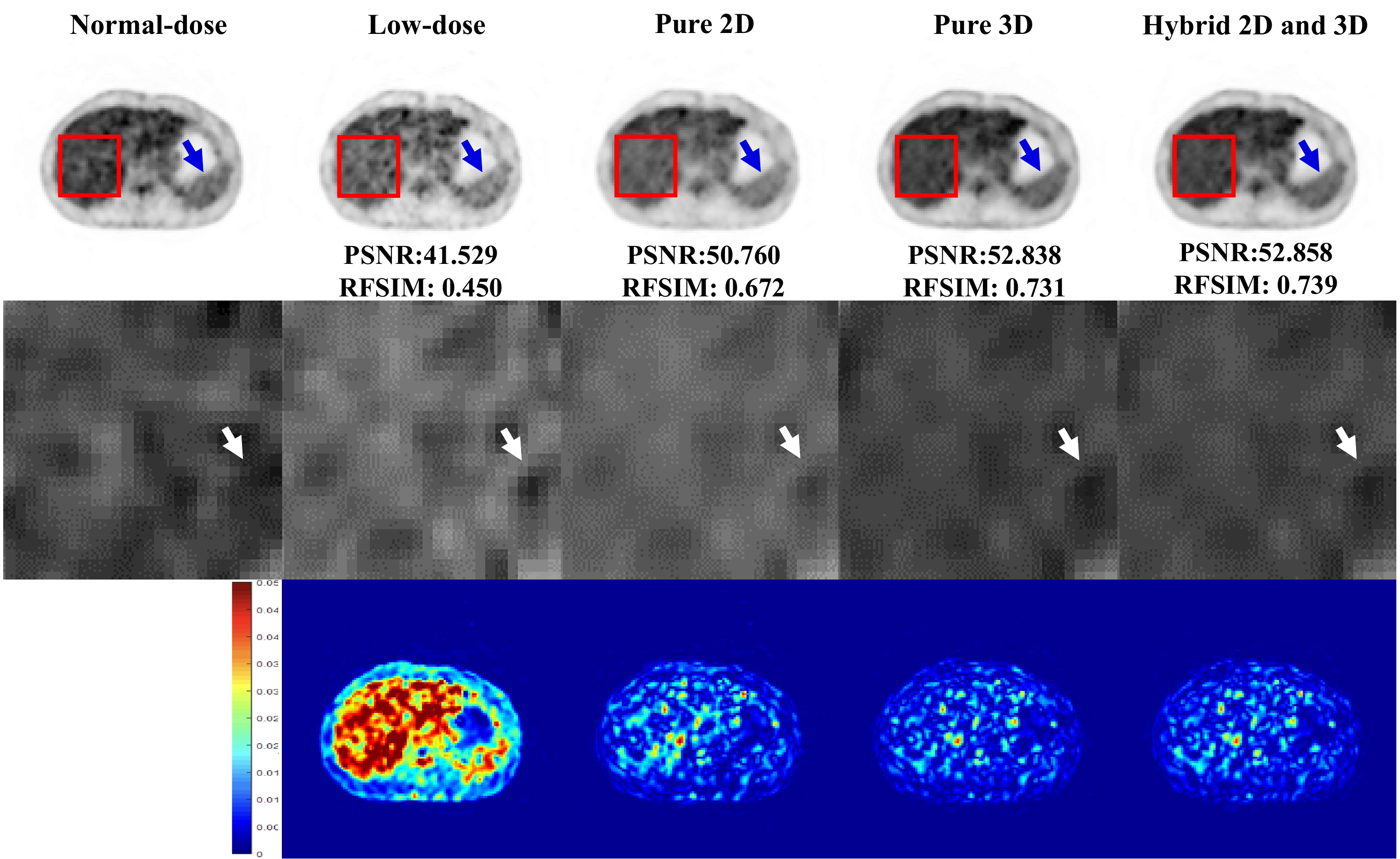}
\caption{Denoising results include the original image, the zoomed ROI marked by the red rectangle on the liver image, and the error map of the pure 2D network, the pure 3D network, and the hybrid 2D and 3D network. The blue arrow indicates the spleen. The white arrow indicates the inferior vena cava. PSNR and RFSIM are shown in the corresponding image. From left to right: normal-dose image; low-dose image; the pure 2D network; the pure 3D network; and the hybrid 2D and 3D network.}
\label{Fig.3}
\end{figure}

Fig. \ref{Fig.3} shows an abdominal section for qualitative comparison. As shown in Fig. \ref{Fig.3}, there is strong image noise in the low-dose PET image. Compared to the desired normal-dose PET image, some regions, for example, the liver and spleen, of the low-dose PET image are fuzzier. It suggests that, with a reduced dose, the reconstructed low-dose PET image has inferior visibility compared to the normal-dose PET image. In reference to the error map shown in Fig. \ref{Fig.3}, the denoising ability of the pure 3D network and the hybrid 2D and 3D network was stronger than that of the pure 2D network. Commonly, the normal liver and spleen are exhibit a heterogeneous mild uptake that contains important texture information. As we can see from the spleen marked by the blue arrow in Fig. \ref{Fig.3}, the pure 3D network and the hybrid 2D and 3D network exhibit an uptake level of the spleen in the denoised image that is closer to the normal-dose image than that of the pure 2D network. The denoising results of the pure 3D network and the hybrid 2D and 3D network recovered more uptake information. These findings indicate the 3D information can boost the low-dose PET image denoising since the number of parameters of the pure 3D network are equal to that of the pure 2D network. The proposed hybrid 2D and 3D network achieved comparable performance to that of the pure 3D network. The reduction of 50\% in the number of parameters in the proposed structure decreased the computational cost significantly. Nevertheless, the denoising performance of the hybrid 2D and 3D network was not ideal since it oversmoothed the low-dose image and led to a loss of texture information. This finding is attributed to the cost of the use of the MSE-based objective function. A region of interest (ROI) was marked by a red rectangle on the liver. When focused on the ROI, oversmoothing reduces the amount of clinically relevant information provided by the denoised image shown in Fig. \ref{Fig.3}. Therefore, it is necessary to find a more suitable training loss function for the low-dose PET image denoising network.

\subsubsection{Training loss function}
Given that the normal-dose PET image contains noise, its use as the target may make the GAN implicitly attempt to include a normal-dose level of noise in the denoised image. Furthermore, only leveraging the adversarial loss cannot guarantee the pixelwise similarity between the two image volumes. In this experiment, we used the identical data and three loss functions, including the MSE loss, adversarial loss, and mixed adversarial and MSE loss, to train the network to verify the effect caused by the training loss function. 

The quantitative results of the denoised image on the testing set and the details of the corresponding network are shown in Table \ref{Table.2}. The network with the 2D and 3D hybrid structure trained by only the MSE loss is labeled as the ‘Hybrid 2D and 3D network’ in Table \ref{Table.2}. The network with the 2D and 3D hybrid structure trained by only the adversarial loss is labeled as ‘WGAN’ in Table \ref{Table.2}. The network with the 2D and 3D hybrid structure trained by the mixed adversarial loss and MSE loss is labeled ‘WGAN (MSE)’. The hybrid 2D and 3D network achieved the highest PSNR and the lowest NRMSE as a result of the MSE-based objective function. As described above, however, it did not guarantee that the denoised images have the best visual perception and texture similarity compared with the normal-dose images. Therefore, the denoising ability of WGAN was not ideal from the perspective of the quantitative analysis. The reason for it was the sole use of the adversarial loss. However, introducing the MSE loss term into the training loss function yielded denoising results in the quantitative evaluation of WGAN (MSE) that were much better than that of WGAN. The metrics related to the human visual perception of the denoising results of WGAN (MSE) were better than that of the hybrid 2D and 3D network, indicating that the introduction of the adversarial loss could improve the visual quality of the denoised image. The combination of the adversarial loss with the MSE loss was a feasible tradeoff between the denoising effect and visual quality.

\begin{figure}[!htbp]
\centering
\includegraphics[width=1\linewidth]{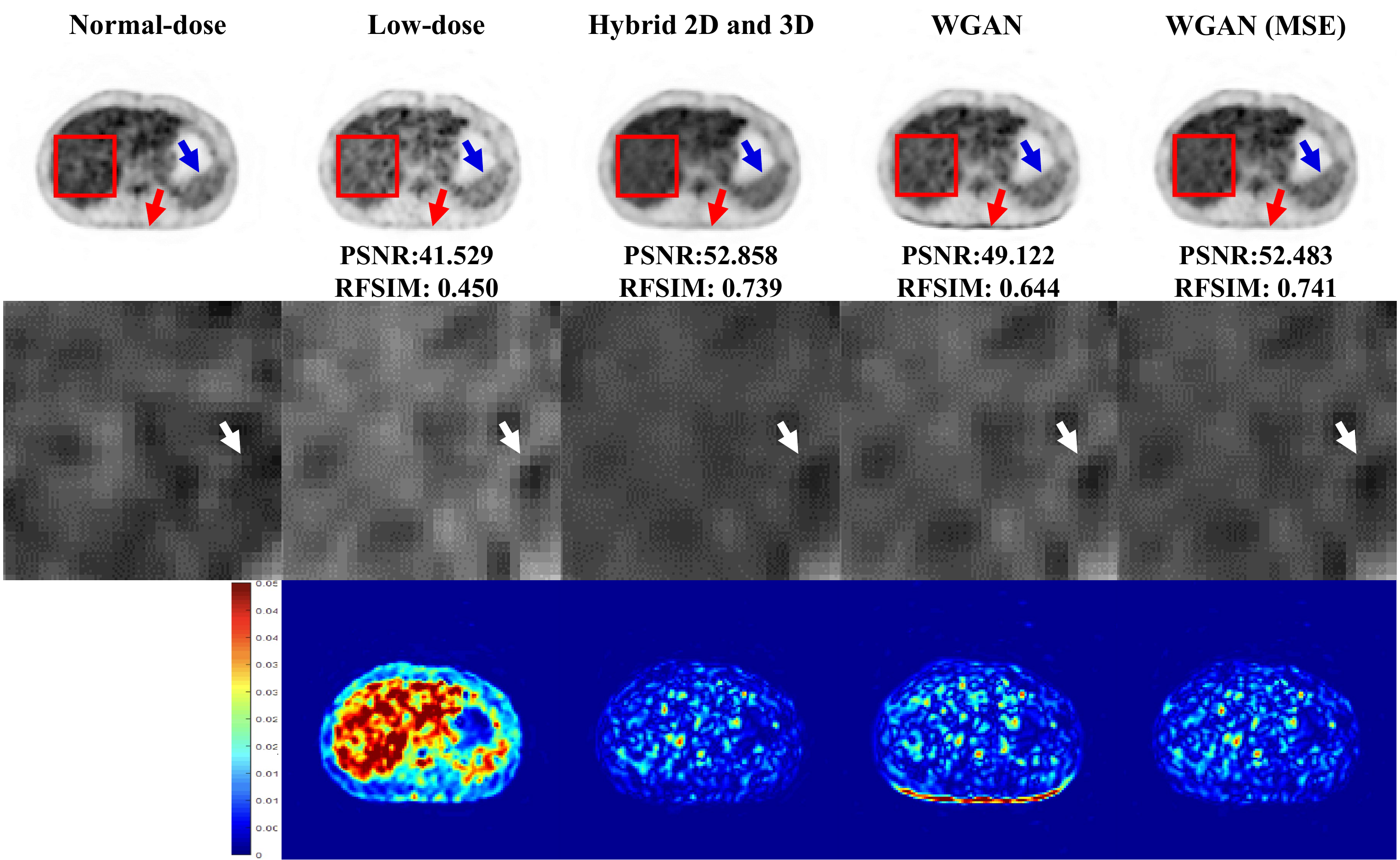}
\caption{Denoising results include the original image, the zoomed ROI marked by the red rectangle on the liver image, and the error map of the network trained by only the MSE loss (Hybrid 2D and 3D network), the network trained by only the adversarial loss (WGAN), and the network trained by the mixed adversarial loss and MSE loss (WGAN (MSE)). The blue arrow indicates the spleen. The white arrow indicates the inferior vena cava. PSNR and RFSIM are shown in the corresponding image. From left to right: normal-dose image; low-dose image; Hybrid 2D and 3D network; WGAN; and WGAN (MSE).}
\label{Fig.4}
\end{figure}

Fig. \ref{Fig.4} shows an abdominal section for qualitative comparison. The hybrid 2D and 3D network produced a denoising result with a lower noise level than that of the WGAN and WGAN (MSE). Nevertheless, the denoising results of WGAN and WGAN (MSE) were substantially better based on visual observation than that of the hybrid 2D and 3D network due to the use of the MSE loss. This finding is attributed to the manner in which MSE guides the mapping from a low-dose image to the normal-dose counterpart, and at the same time smooths a low-dose image to blur noise and details. 
%Editor: Please ensure that the intended meaning has been maintained in this edit.
However, it may oversmooth the resultant image. The zoomed ROI in Fig. \ref{Fig.4} shows that oversmoothing seriously erased the tissue texture in the image denoised by the hybrid 2D and 3D network. In contrast, the results denoised by WGAN and WGAN (MSE) kept more tissue textures since the adversarial loss mitigates the problem by forcing the denoised image to have a distribution of visual characteristics (for example, sharpness) similar to that of the normal-dose image. 

Furthermore, combining the adversarial loss with MSE loss suppressed the generated artifacts effectively. As the skin marked by the red arrow in Fig. \ref{Fig.4} shows, an obvious artifact can be observed in the denoising result of WGAN. It was suppressed in the denoising result of WGAN (MSE) significantly. In summary, the combination of the adversarial loss and MSE loss was a more suitable choice for low-dose PET image denoising.

\subsubsection{Training method}

\begin{table*}[bp]
\caption{Details and the quantitative metrics of denoising results of the models involved in the comparison experiment. For each metric, we mark the best in bold.}
\label{Table.3}
\centering
\setlength{\tabcolsep}{1mm}
{
\begin{tabular}{ccccccccc}
\toprule
& Structure & Loss Function & Training Method & PSNR & NRMSE (\%) & RFSIM & VIF \\ \midrule
Low-dose & - & - & - & 46.148 & 2.916 & 0.541 & 0.555 \\
RED-CNN & 2D & $\mathcal{L}_{\mathrm{MSE}}$ & Direct & 51.922 & 1.503 & 0.656 & 0.656 \\
DCNN & 2D & $\mathcal{L}_{1}$ & Direct & 50.328 & 1.770 & 0.654 & 0.675 \\
CPCE-3D & 2D\&3D & $\mathcal{L}_{\mathrm{adv}}+\mathcal{L}_{\mathrm{per}}$ & Transfer & 54.253 & 1.147 & 0.754 & 0.724 \\
PT-WGAN (MSE) & 2D\&3D & $\mathcal{L}_{\mathrm{adv}}+\mathcal{L}_{\mathrm{MSE}}$ & Transfer & \textbf{55.048} & \textbf{1.048} & \textbf{0.776} & \textbf{0.730} \\ \bottomrule
\end{tabular}
}
\end{table*}
Although GAN is a powerful tool for generating high-quality denoised images, the extensive computational cost and unstable convergence characteristic raise the its threshold. This experiment was designed to verify the effect of the proposed task-specific initialization based on transfer learning. 

In this experiment, we used three loss functions, including the MSE, SSIM, and perceptual loss, to separately train the generator of PT-WGAN first. Then, we used the proposed task-specific initialization to initialize the generator. The network directly trained by the mixed adversarial and MSE loss is labeled as ‘WGAN (MSE)’ in Table \ref{Table.2}. The networks transferred from the pretrained model trained by the MSE loss, SSIM loss, and perceptual loss are labeled as ‘PT-WGAN (MSE)’, ‘PT-WGAN (SSIM)’, and ‘PT-WGAN (Perceptual)’, respectively, in Table \ref{Table.2}. Among them, PT-WGAN (MSE) was transferred from the abovementioned hybrid 2D and 3D network. The quantitative analysis results of the denoised image on the testing set and the details of the corresponding network are shown in Table \ref{Table.2}. In reference to the quantitative analysis metrics shown in Table \ref{Table.2}, the denoising performance of PT-WGAN (MSE) was slightly better than that of WGAN (MSE).

\begin{figure}[!hbtp]
\centering
\includegraphics[width=1\linewidth]{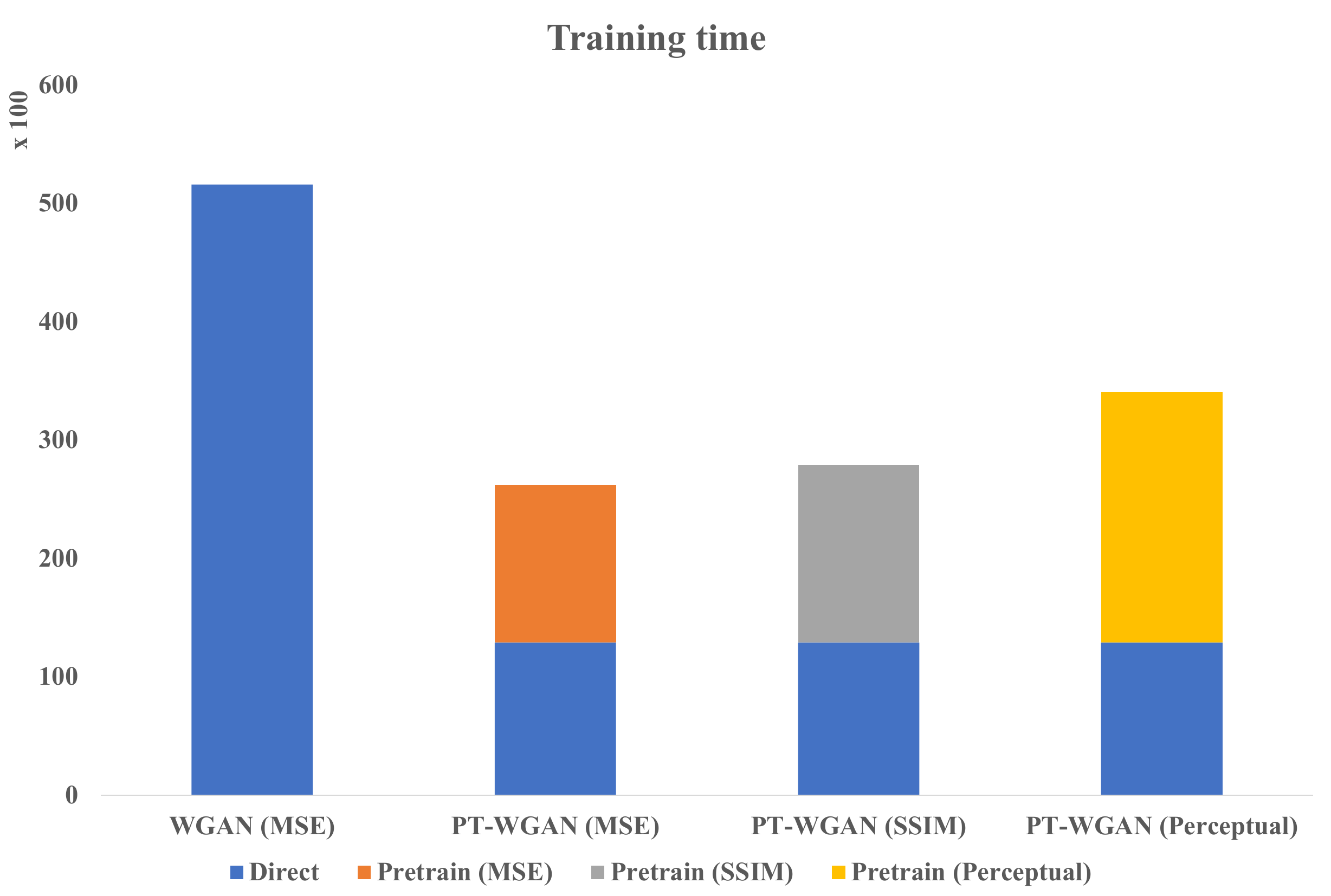}
\caption{Training time of WGAN (MSE), PT-WGAN (MSE), PT-WGAN (SSIM), and PT-WGAN (Perceptual). Let ``Direct'' denote the training time of a network trained by the mixed adversarial and MSE loss, and ``Pretrain (MSE), Pretrain (SSIM), Pretrain (Perceptual)'' denote the training time of a network pretrained by the MSE loss, SSIM loss, and perceptual loss, respectively.}
\label{Fig.5}
\end{figure}

In this study, each pretrained model took 30 epochs to converge. The generator pretrained by the MSE loss, SSIM loss, and perceptual loss required 445 s, 500 s, and 705 s per epoch, respectively. However, the generator trained directly in the WGAN framework required 1,290 s to finish one epoch. The network that leveraged the task-specific initialization only required approximately 10 epochs to achieve the same denoising performance as that of the network trained from scratch after 40 epochs. As shown in Fig. \ref{Fig.5}, training the network via task-specific initialization based on the network pretrained by the MSE loss can save 50\% of the computational time in this case. In addition, it effectively avoids the divergence problem in the training process of GAN. In our experiments, we trained WGAN (MSE), PT-WGAN (MSE), PT-WGAN (SSIM), and PT-WGAN (Perceptual) ten times. Among them, WGAN (MSE) failed to converge four times. PT-WGAN (MSE), PT-WGAN (SSIM), and PT-WGAN (Perceptual) did not diverge. The best results are in Table \ref{Table.2}. The proposed task-specific initialization helps obtain a satisfactory GAN model. Considering the computational efficiency and the denoising effect, we selected PT-WGAN(MSE) as the best model in this study.

\subsection{Comparison Experiment}
\begin{figure*}[!hbtp]
\centering
\includegraphics[width=1\linewidth]{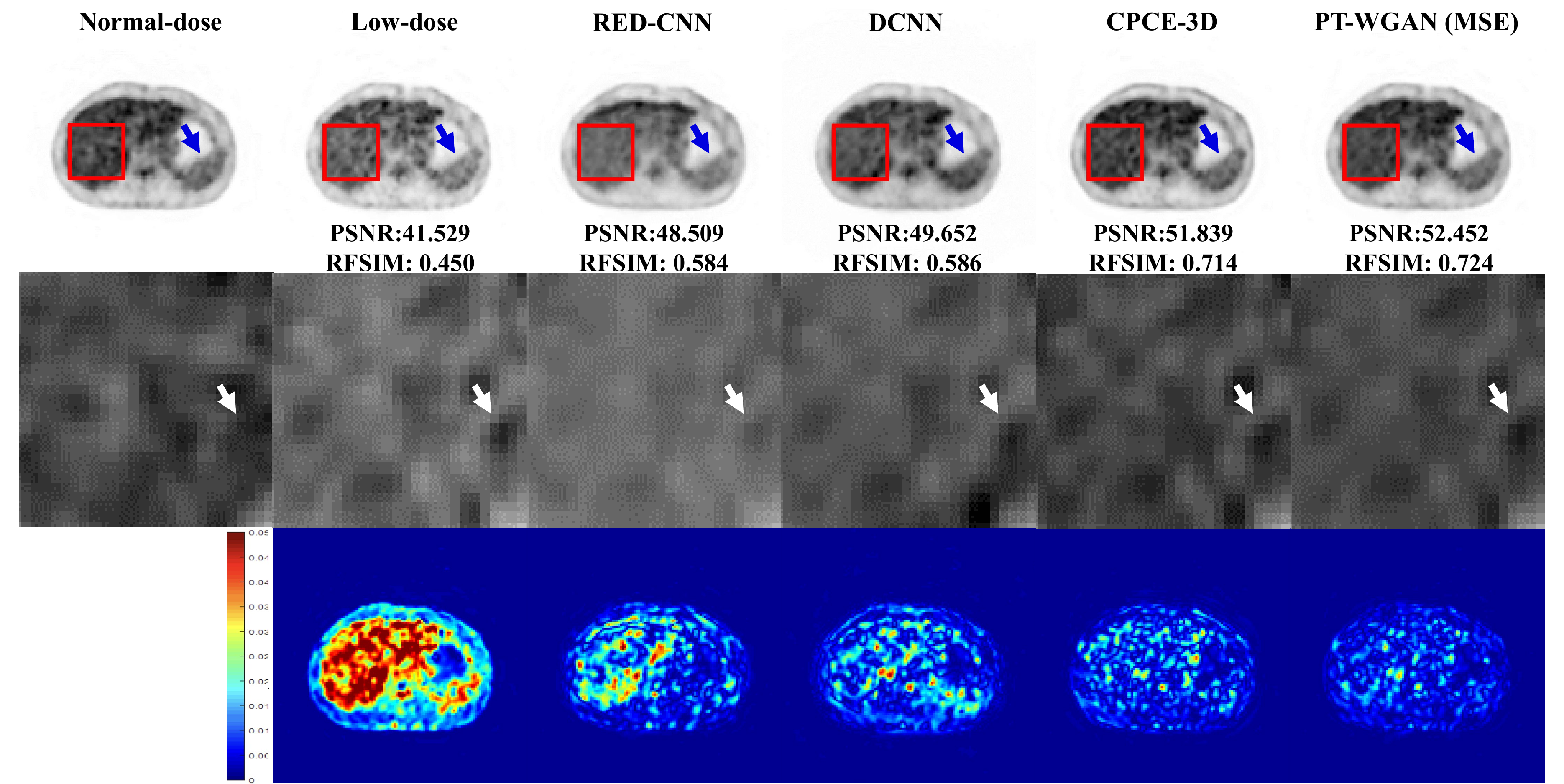}
\caption{Denoising results include the original image, the zoomed ROI marked by the red rectangle on the liver image, and the error map image of different networks. The blue arrow indicates the spleen. The white arrow indicates the inferior vena cava. PSNR and RFSIM are shown in the corresponding image. From left to right: normal-dose image; low-dose image; RED-CNN; DCNN; CPCE-3D; and PT-WGAN (MSE)}
\label{Fig.6}
\end{figure*}

To demonstrate the denoising performance of the proposed PT-WGAN (MSE), we also trained three recently published state-of-the-art methods include RED-CNN \cite{chen2017low}, DCNN \cite{xu2017200x}, and CPCE-3D \cite{shan20183}.

The RED-CNN proposed by Chen \etal \cite{chen2017low} is a CNN. For the network structure, it leverages the combination of 2D convolution and deconvolution operators and introduces additional shortcut connections to avoid gradient diffusion while helping the network recover details. The training loss function is MSE loss. The training method employs training from scratch.

The DCNN proposed by Xu \etal \cite{xu2017200x} is also a CNN. For the network structure, it uses a modified U-net structure. Shallow and deep features are concatenated to improve the fitting ability of the DCNN. The DCNN is the only network of the three considered networks that uses pooling layers	 since the number of convolution kernels used is large. The training loss function is the $L_{1}$-norm. Similar to the RED-CNN, it is also trained from scratch.

The CPCE-3D proposed by Shan \etal \cite{shan20183} is a WGAN. For the network structure, it expands the image denoising from the 2D domain to the 3D domain. The CPCE-3D is also a hybrid 2D and 3D network. Similar to the DCNN, shallow and deep features are concatenated together in the CPCE-3D. The difference is that the CPCE-3D introduces external 2D convolution layers with a $1 \times 1$ kernel after feature concatenation for feature dimension reduction. The training loss function is implemented with mixed adversarial and perceptual loss functions. For the training method, it is trained by initializing the parameters of the CPCE-3D from a pretrained CPCE-2D network.

\begin{figure*}[!hbtp]
\centering
\includegraphics[width=1\linewidth]{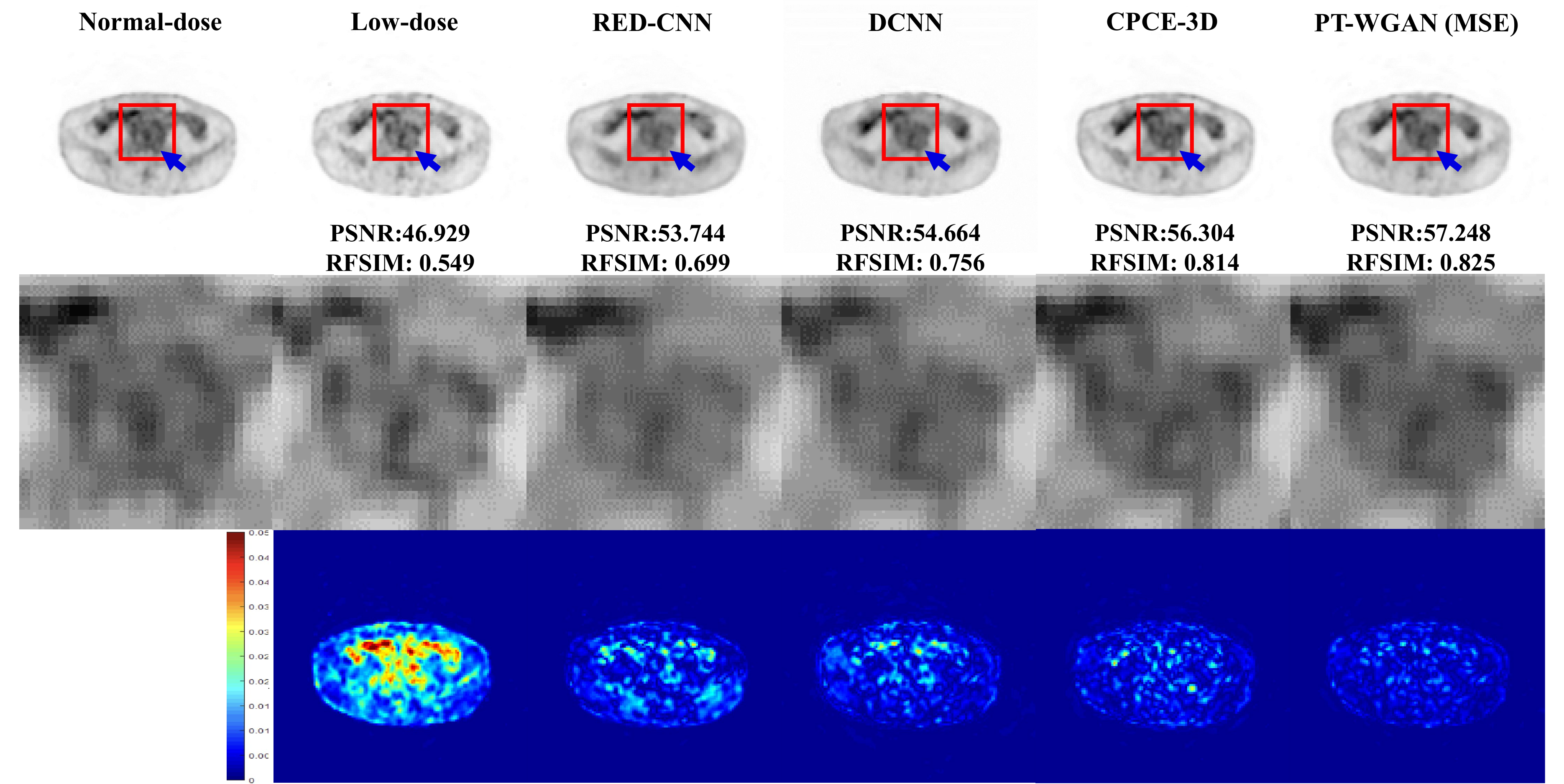}
\caption{Denoising results include the original image, the zoomed ROI marked by the red rectangle on the vertebral body image, and the error map image of different networks. The blue arrow indicates the vertebral body. PSNR and RFSIM are shown in the corresponding image. From left to right: normal-dose image; low-dose image; RED-CNN; DCNN; CPCE-3D; and PT-WGAN (MSE)}
\label{Fig.7}
\end{figure*}

The quantitative comparison of the denoising performance of different networks is shown in Table \ref{Table.3}. As the abdomen and bone are common locations of various primary and metastatic tumors, abdominal and skeletal PET scans are of extreme importance in clinical practice. Therefore, we selected two slices consisting of abdominal parenchymal organs (liver and spleen), and bone (vertebral body) to show the denoising performance of different networks, which are shown in Figs. \ref{Fig.6} and \ref{Fig.7}. Composed of rich blood sinus, the normal liver and spleen frequently exhibit a heterogeneous mild uptake. Similarly, the vertebral body contains abundant trabecular tissues that divide the marrow tissue and typically have a heterogeneous mild uptake. In our study, we implement the denoising method to recover the organ uptake information while providing pixelwise texture similarity with the normal-dose image.

As we can see in Table \ref{Table.3}, each metric of the denoising results of the RED-CNN and DCNN was better than that of the low-dose image. It means that the RED-CNN and DCNN approaches significantly improved the quality of the low-dose image. However, the RED-CNN did not recover enough organ uptake information. As we can see in the spleen marked by the blue arrow in Fig. \ref{Fig.6}, there is still a large difference compared to the normal-dose image. Meanwhile, the RED-CNN seriously erased the tissue texture, as shown in the zoomed ROI on the liver in Fig. \ref{Fig.6}. Although the DCNN recovered more organ uptake information than did the RED-CNN, oversmoothing is still observed in the denoising results. As revealed in the zoomed ROI images in Figs. \ref{Fig.6} and \ref{Fig.7}, many tissue textures were lost in the denoising results of the DCNN.

Compared with the RED-CNN and DCNN, the CPCE-3D obtained better quantitative metrics, especially regarding RFSIM and VIF, indicating that the denoised images of CPCE-3D had a better visual performance. In reference to the error map shown in Figs. \ref{Fig.6} and \ref{Fig.7}, the CPCE-3D approach suppressed more noise than did the RED-CNN and DCNN approaches. In regard to the vertebral body marked by the blue arrow shown in Fig. \ref{Fig.7}, the uptake information shown in the denoised image of CPCE-3D is relatively closer to that shown in the normal-dose image among these network comparisons. 
%Editor: Please ensure that the intended meaning has been maintained in this edit.
Meanwhile, in reference to the zoomed ROI image on the vertebral body, many tissue textures were retained in the denoising process of the CPCE-3D approach.

The denoising effect of the PT-WGAN (MSE) exhibits a certain improvement compared with that of the CPCE-3D. In terms of the four quantitative metrics, namely, PSNR, NRMSE, RFSIM, and VIF, our network achieved the best results, as shown in Table \ref{Table.3}. After being processed by the PT-WGAN (MSE), the low-dose images demonstrated approximately the same metabolic activity of abdominal parenchymal organs (for example, the liver and spleen), and bone (for example, the vertebral body) as the normal-dose images, as shown in Figs. \ref{Fig.6} and \ref{Fig.7}. When ROIs were focused on, there are more tissue textures in the denoised images of the PT-WGAN (MSE) than that of the other networks, as shown in Figs. \ref{Fig.6} and \ref{Fig.7}. In summary, the PT-WGAN (MSE) greatly reduced the noise while maintaining the nearest approximation of visual image quality to the normal-dose image. Hence, the proposed PT-WGAN (MSE) provides the potential for evaluating the normal metabolism of the human body with an 80\% radiotracer dose reduction.

\section{DISCUSSIONS AND CONCLUSION}
The novel features of the proposed network include the well-designed structure aided by the noise reduction loss term and the task-specific initialization based on transfer learning.

It is important to reduce the loss of details while denoising low-dose PET image volumes. We introduced 3D deconvolutional operators to recover the details discarded by 3D convolution. Then, we introduced skip connections between the corresponding layers to make the best tradeoff between noise suppression and detail preservation. There were not only 3D but also 2D convolutional operators. It is practical to use both 3D contextual and planar information for low-dose PET image denoising. The use of hybrid 2D and 3D convolutional operators is beneficial for saving computational resources while maintaining the denoising performance.

Generating a normal-dose image volume from a low-dose image volume is the ultimate goal. However, achieving it perfectly is impossible; therefore, in practice, we only attempt to generate the image volume that is as close to the normal-dose image volume as possible. The denoising results of \cite{shan20183} achieved excellent quality by taking advantage of a GAN. However, the authenticity of the images generated by GANs can be problematic if not well constrained. Furthermore, the training of a GAN is very challenging. To address these issues, we proposed a task-specific initialization based on transfer learning coupled with the MSE loss, and our approach was shown to demonstrate success.

For task-specific initialization, we used the parameters of a pretrained CNN model to initialize the parameters of the generator of the PT-WGAN. The reasonable starting point transferred from the pretrained model enabled a well-informed training process of the PT-WGAN. The parameters of the pretrained network were proven to be a good starting point that was close to the optimal point for the PT-WGAN and made the optimization more feasible. While the training difficulty of a CNN is low, that of a GAN is high. The task-specific initialization based on a transfer learning strategy overcomes the training difficulty of GANs. Based on the common characteristics of underlying images, we transferred a pretrained CNN model to a GAN, which made the training process of the GAN efficient. This kind of task-specific initialization is an efficient and heuristic approach. Notably, the performance of the proposed network with knowledge transferred from the pretrained network was better than that of the proposed network trained directly. The experimental results described above have demonstrated that the use of task-specific initialization can simultaneously reduce the training difficulty and improve the performance and the training efficiency of the proposed network.

We believe that the effect of network-based image denoising is related to the width and depth. How to optimize the network topology is a question worthy of further study. With the development of GANs, an increasing number of well-structured GAN networks have emerged. Among them, WGAN is not necessarily the best. Therefore, there is still considerable room for the development of denoising methods for low-dose PET imaging that can be achieved by developing more advanced networks than what was proposed here. Meanwhile, a reversible neural network needs to be developed since mapping the normal-dose PET image from the corresponding low-dose image is an ill-posed problem. The reversible neural network may provide a better solution. Finally, the performance of supervised PET image denoising via deep learning critically depends on the quantity of labeled data, and self-supervised learning can help further. The use of self-supervised learning for PET image denoising will be our next step.

In conclusion, we proposed a parameter-transferred Wasserstein generative adversarial network (PT-WGAN) for low-dose PET image denoising. The experimental results on clinical data show that the proposed network can suppress image noise more effectively while preserving better image fidelity than three selected state-of-the-art methods. Further work is in progress to improve the performance of low-dose PET denoising.

\section{ACKNOWLEDGMENTS}
The authors would like to thank the anonymous reviewers and associate editor for their constructive comments and suggestions and also wish to express gratitude to Neusoft Medical Systems Co., Ltd and the patients who provided data. In addition, Mr. Gong would like to thank Huimin Zheng for her great support and patience during this period.

\end{document}